\documentclass[aps,prb,twocolumn]{revtex4}
\usepackage{epsfig}
\usepackage[dvipsnames,usenames]{color}
\usepackage{color}
\usepackage{amsmath}
\usepackage{amssymb}
\usepackage{hyperref}
\usepackage{graphicx}

\usepackage{wasysym}
\usepackage{times}
\usepackage{comment}
\usepackage{array}
\usepackage{multirow}
\usepackage{tabularx}
\usepackage{float}
\usepackage[utf8]{inputenc}
\usepackage[T1]{fontenc}
\usepackage{cancel}
\usepackage{ulem}

\emergencystretch=\maxdimen
\newcommand{\bK}{{\bf K}}

\begin{document}
\title{Characterizing the superconducting instability in a two-orbital $d$-$s$ model: \\insights to infinite-layer nickelate superconductors}
\author{Mi Jiang}
\affiliation{Institute of Theoretical and Applied Physics, Jiangsu Key Laboratory of Thin Films, School of Physical Science and Technology, Soochow University, Suzhou 215006, China}

\begin{abstract}
Motivated by the recent realization of the infinite-layer nickelate superconductivity (SC), we quantitatively investigate a two-orbital $d$-$s$ model by dynamic cluster quantum Monte Carlo calculations. Focusing on the impact of inter-orbital hybridization on the superconducting properties, our simulations indicate that the $d$-$s$ hybridization strength has a decisive role in the suppression of the $d$-wave pairing in the doping regime relevant to infinite-layer nickelates. Although we confirm on the single-orbital description at weak hybridization, at relatively large hybridization, there exists a constructive effect of the non-negligibly finite $s$ orbital occupancy on inhibiting the suppression of the superconductivity. More strikingly, there exists a possible SC enhancement at large enough Hubbard interaction via large $d$-$s$ hybridization.
We further provide some insights on the relevance of $d$-$s$ model to infinite-layer nickelates.
\end{abstract}

\maketitle

\section{Introduction}

The realization of the superconductivity (SC) in Sr-doped NdNiO$_2$ films grown on a SrTiO$_3$ substrate~\cite{2019Nature} is the breakthrough in the decade-long pursuit of rare-earth nickelates~\cite{Junjie_review,Botana_review}. Even more fascinating is the successful synthesis of superconducting doped Pr$_{1-x}$Sr$_{x}$NiO$_2$~\cite{Pr}, La$_{1-x}$Ca$_{x}$NiO$_2$~\cite{LaSC}, and undoped quintuple-layer Nd$_6$Ni$_5$O$_{12}$~\cite{Nd6Ni5O12}, which indicates the advent of Ni-based superconducting family. Undoubtedly, the most important question is the pairing symmetry and mechanism in this new family of superconductors and furthermore its similarity and difference from other transition metal based superconductors like Cu- and Fe-based compounds. For example, given that the parent compound of infinite-layer NdNiO$_2$ is discovered not to host the long-range magnetic order~\cite{Pickett2004,Hepting,dft1,dft9,dft16,dft18,dft26,dft25,noLRO} in spite of magnetic correlations~\cite{mag1,mag2,mag3,mag4,mag5}, whether the debated spin-fluctuation scenario for cuprate superconductors also applies for the infinite-layer nickelates is an intriguing question. 

Physically, the minimal microscopic model to capture the essential physics of infinite-layer nickelates is a demanding task. Previous numerous electronic structure calculations based on DFT~\cite{Pickett2004,dft1,dft2,dft3,dft4,dft5,dft6,dft7,dft8,dft9,dft10,dft24,dft26} and/or DFT+DMFT~\cite{dft101,dft102,dft11,dft12,dft13,dft14,dft15,dft16,dft17,dft18,dft19,dft20,dft21,dft22,dft23,dft25,Held2020,Held2022,Hanghuireview} methodologies have provided fruitful understanding of the electronic structure of infinite-layer nickelates including R$_{n+1}$Ni$_{n}$O$_{2n+2}$ with R = La, Nd, Pr for varying $n$~\cite{Botana_review,Hanghuireview,Held2022}, whose calculated properties have been partly confirmed by recent experiments for $n=5$~\cite{Nd6Ni5O12}. One distinct feature associated with infinite-layer nickelates lies in the presence of the Nd-derived bands, whose self-doping effect on the magnetic and superconducting properties is highly debated, which is manifested by the opposite arguments supporting the single Ni-$d_{x^2-y^2}$ band description~\cite{oneband,dft101,dft4,dft24} versus the multi-orbital scenarios~\cite{Hepting}, e.g. Ni-$e_g$ with or without other Nd-5d orbitals. 

The most significant theoretical reasoning underlying the single-orbital scenario arises from the demonstration of the depletion of the rare-earth bands, namely the extremely low electron density of rare-earth layer, which results in a Fermi surface reconstruction that might render the infinite-layer nickelates essentially akin to cuprates. Despite there have been strong theoretical~\cite{Held2022,dft101,dft4,dft24} and experimental~\cite{oneband} support on this scenario, it is still not completely resolved whether and how the low occupation of rare-earth bands impact the charge~\cite{Pr,Hall1,Hall2}, magnetic~\cite{GuangMing,GuangMing1}, and superconducting properties~\cite{Mi2020,dswave,dft2,dft3,dft101,dft102,spinfluc1,spinfluc2} in the whole phase diagram.

\section{Model and Methodology}
\subsection{Two-orbital $d$-$s$ model}
To provide more insights on the prominent role of the rare-earth bands, we employ a minimal two-orbital $d$-$s$ model, where $d$-orbital mimics Ni-$3d_{x^2-y^2}$. Additionally, we include the rare-earth R-$5d$ orbitals, which can be effectively approximated as an interstitial s-like orbital located at the missing apical Oxygen above and below Ni in the rare-earth layers. The Hamiltonian reads as follows
\begin{align} \label{eq:HM}
	\hat{H} = & \sum_{k\sigma} (E^d_k n^d_{k\sigma} +E^s_k n^s_{k\sigma}) + U \sum_i n^d_{i\uparrow}n^d_{i\downarrow}\nonumber\\ 
	& + \sum_{k\sigma} V_k (d^\dagger_{k\sigma}s^{\phantom\dagger}_{k\sigma}+h.c.) 
\end{align}
with two orbitals' dispersion and $d$-$s$ hybridization as
\begin{align} \label{dis}
	E^d_k = & -2 t_d (\cos k_x+\cos k_y)+4 t'_d \cos k_x\cos k_y -\mu \nonumber\\ 
	E^s_k = & -2 t_s (\cos k_x+\cos k_y) + \epsilon_s -\mu \nonumber\\ 
	V_k = & -2 V (\cos k_x+\cos k_y)
\end{align}
where $d^{\dagger}_{k\sigma}(s^{\dagger}_{k\sigma})$ are creation operators in momentum space for two orbitals. $n^{d(s)}_{i\sigma}$ and $n^{d(s)}_{k\sigma}$ are the associated number operators in the real and momentum spaces separately. The chemical potential $\mu$ can be tuned for a desired total electron density. The on-site energy $\epsilon_s$ of $s$ orbital is a tunable parameter to adjust the relative electron density between two orbitals with the criterion that the $s$ orbital's occupancy is around 6-10\% in an undoped system (total electron density $n=1$) in order to be consistent with the experimental finding that there are approximately 8\% electrons per formula unit for NdNiO$_2$~\cite{Hall1}.
We adopt the convention that the nearest-neighbor hopping between $d$ orbitals $t_d=1$ is set as the energy unit.  Following the recent numerical investigation~\cite{TPD_ds}, we choose $t'_d=-0.25, t_s=0.5$ as typical hopping integrals. 
Besides, the impact of on-site Coulomb repulsion $U$ of $d$ orbital will be studied in the present work.

The specific dispersion of the $d$-$s$ hybridization $V_k$ might quantitatively affect the detailed evaluation of various physical properties. Nevertheless, as pointed out by Ref.~\cite{TPD_ds}, although the symmetry analysis of the real infinite-layer materials reveals that the inter-layer hoppings between Ni-3d and the rare-earth R-5d orbitals can be negligibly small, the electrons in the rare-earth layer are highly itinerant so that the detailed form of $V_k$ probably only has minor effects. For simplicity, therefore, we focus on the model Hamiltonian where $d$-$s$ has only nearest-neighbor hoppings~\cite{dft25}. 

One important issue we will explore is the magnitude of $V$, which is believed to be small $V/t_d \sim 0.1$ in previous studies~\cite{TPD_ds}. Strikingly, however, a recent investigation indicated that the dominant hybridization between Ni-3d orbitals and itinerant electrons of this interstitial s orbital comes from a large inter-cell hopping~\cite{dft25}. Therefore, the present work will not restrict our attention to only small hybridization $V$ but will study the influence of its magnitude. Specifically, we choose both $V/t_d = 0.1$~\cite{TPD_ds} and $V/t_d = 0.6$ estimated from Ref.~\cite{dft25} as two typical values. The focus in this work is to investigate the relevance of the $d$-$s$ model to the infinite-layer nickelates, in particular the most attractive superconducting instabilities in the hole doped systems. 

Before proceeding, we remark on some extensively investigated limits of the $d$-$s$ model. On the one hand, at $V=0$, namely the two-dimensional single-band Hubbard model, there have been tremendous theoretical effort for its solutions and more importantly its relevance in strongly correlated quantum materials like cuprate and nickelate superconductors. Regarding its superconducting properties, the consensus is that the $d$-wave pairing is appreciated owing to its retarded nature to minimize the impact of local repulsion $U$~\cite{ScalapinoRMP,Maier06}. On the other hand, at finite hybridization $V$ but $t_d=t'_d=0$, the model recovers the conventional periodic Anderson model (PAM), which is widely treated as the standard model of describing the heavy fermion physics. The deeper connection and moreover the transition or crossover between PAM and $d$-$s$ model will be left for another future independent study.

\subsection{Dynamical cluster approximation}
We adopt the dynamical cluster approximation (DCA)~\cite{Hettler98,Maier05,code} with a continuous time auxilary field (CT-AUX) quantum Monte Carlo (QMC) cluster solver~\cite{GullCTAUX} to
numerically solve the $d$-$s$ model Eq.~\eqref{eq:HM}.
As one of the successful many-body methods, DCA has provided much insight on the strongly correlated physics especially for the Hubbard-type models~\cite{Hettler98,Maier05}. In principle, DCA evaluates the physical properties in the thermodynamic limit via mapping the bulk lattice problem onto a finite cluster embedded in a mean-field self-consistently, where the effective cluster problem is exactly solved by various perturbative or non-perturbative many-body methods~\cite{Hettler98,Maier05}. In practice, DCA is realized by a self-consistent loop for the convergence between the cluster and coarse-grained single-particle Green's function. The truly time-consuming part of the DCA loop lies in the cluster solver, e.g. CT-AUX adopted here. In other words, most uncertainties of DCA method arises from the cluster solver method e.g. quantum Monte Carlo procedure. 
To simulate a wide range of doping levels and Hubbard interaction, we stick on the smallest $N_c=2\times2$ DCA cluster to keep the QMC sign problem manageable~\cite{GullCTAUX} 
down to the superconducting transition temperatures $T_c$. 
In fact, the qualitative trend of our major results can be verified with more expensive simulations employing larger clusters such as $N_c=8$. 

To characterize the superconducting instability, we rely on the leading eigenvalues of the Bethe-Salpeter equation (BSE) in the eigen-equation form in the particle-particle channel~\cite{Maier06,Scalapino06}
\begin{align} \label{BSE}
    -\frac{T}{N_c}\sum_{K'}
	\Gamma^{pp}(K,K')
	\bar{\chi}_0^{pp}(K')\phi_\alpha(K') =\lambda_\alpha(T) \phi_\alpha(K)
\end{align}
where $\Gamma^{pp}(K,K')$ denotes the lattice irreducible particle-particle vertex of the effective cluster problem with combining the cluster momenta $\bK$ and Matsubara frequencies $\omega_n=(2n+1)\pi T$ to $K=(\bK,i\omega)$. One of the key DCA assumptions is that the desired lattice two-particle irreducible vertex $\Gamma$ is approximated by the its cluster counterpart $\Gamma_c$, which can be obtained by 
\begin{align} \label{e2}
  \chi_{c\sigma\sigma'}(q,K,K') &= \chi^0_{c\sigma\sigma'}(q,K,K') + \chi^0_{c\sigma\sigma''}(q,K,K'') \nonumber \\
  & \times \Gamma_{c\sigma''\sigma'''}(q,K'',K''') \chi_{c\sigma'''\sigma'}(q,K''',K')
\end{align}
where the cluster two-particle Green's function 
\begin{align} \label{e1}
  \chi_{c\sigma\sigma'}(q,K,K') &= \int^{\beta}_0 \int^{\beta}_0 \int^{\beta}_0 \int^{\beta}_0 d\tau_1 d\tau_2 d\tau_3 d\tau_4 \nonumber \\
  & \times e^{i[(\omega_n+\nu)\tau_1 -\omega_n\tau_2 +\omega_{n'}\tau_3 -(\omega_{n'}+\nu)\tau_4]} \nonumber \\
  \times \langle \mathcal{T} & c^{\dagger}_{K+q,\sigma}(\tau_1) c^{\phantom{\dagger}}_{K\sigma}(\tau_2) c^{\dagger}_{K'\sigma'}(\tau_3) c^{\phantom{\dagger}}_{K'+q,\sigma'}(\tau_4) \rangle
\end{align}
with conventional notation $K=(\mathbf{K},i\omega_n)$, $K'=(\mathbf{K'},i\omega_{n'})$, $q=(\mathbf{q},i\nu)$ and the time-ordering operator $\mathcal{T}$ can be calculated numerically via a DCA cluster solver (CT-AUX in our case). Concurrently, the non-interacting two-particle Green's function $\chi^0_{c\sigma\sigma'}(q,K,K')$ is constructed from the product of a pair of fully dressed single-particle Green's functions. Note the usual convention that the summation is to be made for repeated indices. In this work, we are mostly interested in the even-frequency even-parity (spin singlet) $d$-wave pairing tendency so that $q=(\mathbf{q},i\nu)=0$ is assumed~\cite{Maier06,Scalapino06}, despite that there exists odd-frequency pairing channels at high temperatures that quickly decays with lowering $T$.

In the BSE Eq.~\eqref{BSE}, the coarse-grained bare particle-particle susceptibility
\begin{align}\label{eq:chipp}
	\bar{\chi}^{pp}_0(K) = \frac{N_c}{N}\sum_{k'}G(K+k')G(-K-k')
\end{align}
is obtained via the dressed single-particle Green's function $G(k)\equiv G({\bf k},i\omega_n) =
[i\omega_n+\mu-\varepsilon_{\bf k}-\Sigma({\bf K},i\omega_n)]^{-1}$, where $\mathbf{k}$ belongs to the DCA patch surrounding the cluster momentum $\mathbf{K}$, $\mu$ the chemical potential, $\varepsilon_{\bf k}$ the
dispersion relation, and $\Sigma({\bf K},i\omega_n)$ the cluster self-energy. 

In practice, we normally choose 16 or more discrete points for both the positive and negative fermionic Matsubara frequency $\omega_n=(2n+1)\pi T$ mesh for measuring the two-particle Green's functions and irreducible vertices $\Gamma$. Therefore, the BSE Eq.~\eqref{BSE} effectively becomes an eigenvalue problem of a matrix of size $(32N_c)\times (32N_c)$, which can be routinely solved straightforwardly. 

Physically, the magnitude of the eigenvalue $\lambda_\alpha(T)$ denotes the normal state pairing tendency in the superconducting channel. Correspondingly, the spatial, frequency, and orbital dependence of the eigenvector $\phi_\alpha({\bf K},i\omega_n)$ can be viewed as the normal state analog of the superconducting gap function to reflect the structure of the pairing interaction $\Gamma^{pp}$~\cite{Maier06,Scalapino06}.
The superconducting $T_c$ is extracted via the temperature where the leading eigenvalue of Eq.~\eqref{BSE} $\lambda(T_c)=1$.
Throughout this work, it is found that the leading pairing symmetry occurs for the $d$-wave channel with momentum structure $\cos K_x - \cos K_y$ so that we are only concerned in the leading eigenvalues $\lambda_d$ and associated $\phi_d({\bf K},i\omega_n)$. In other words, the presence of $s$ orbital and its hybridization with $d$ orbital does not affect the $d$-wave pairing symmetry of the conventional Hubbard model with solely $d$ orbital. Its implication on the infinite-layer nickelates including the recent experimental findings will be discussed later on.

\section{Results}
\begin{figure}
\psfig{figure=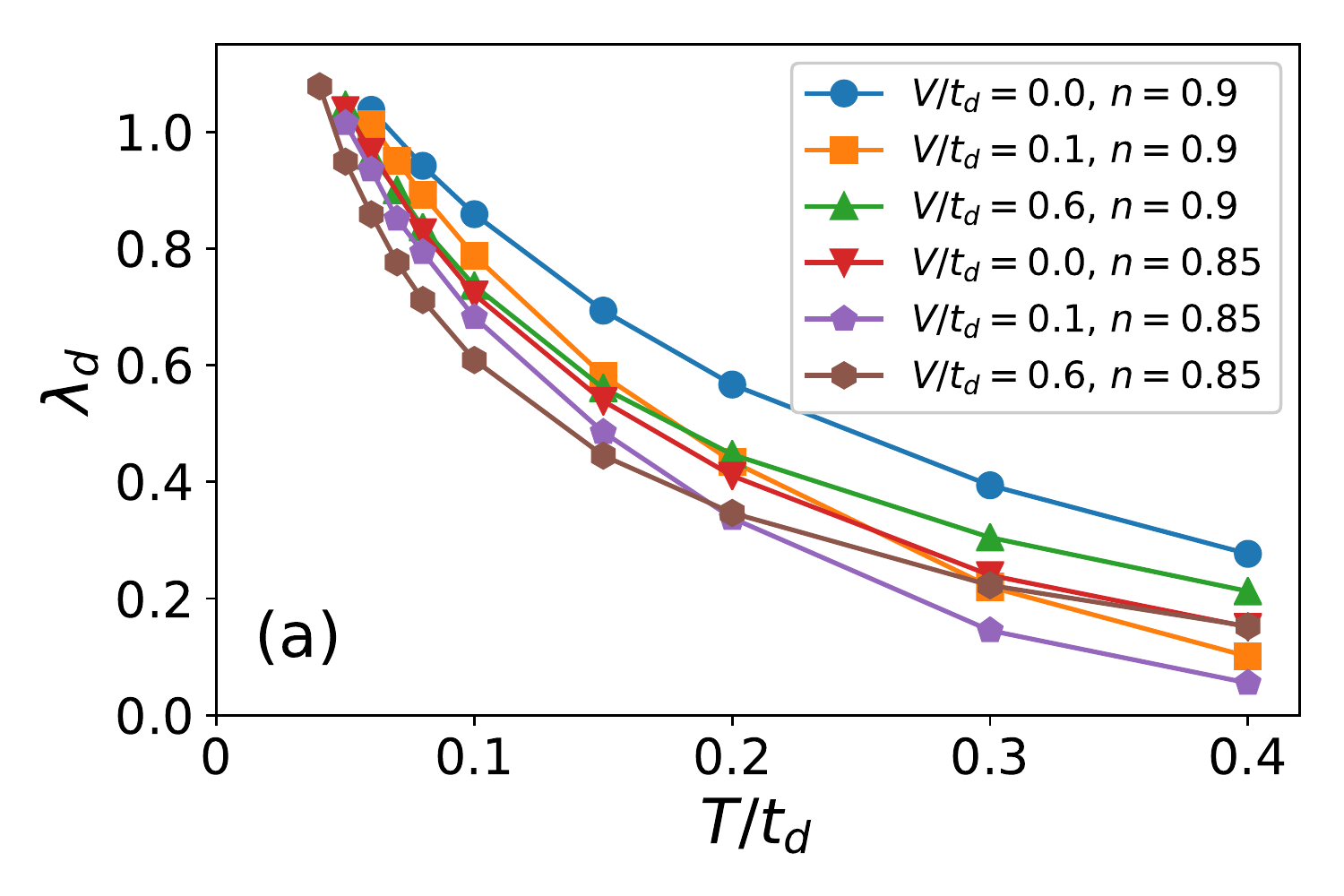,height=5.5cm,width=8.5cm,angle=0,clip} 
\psfig{figure=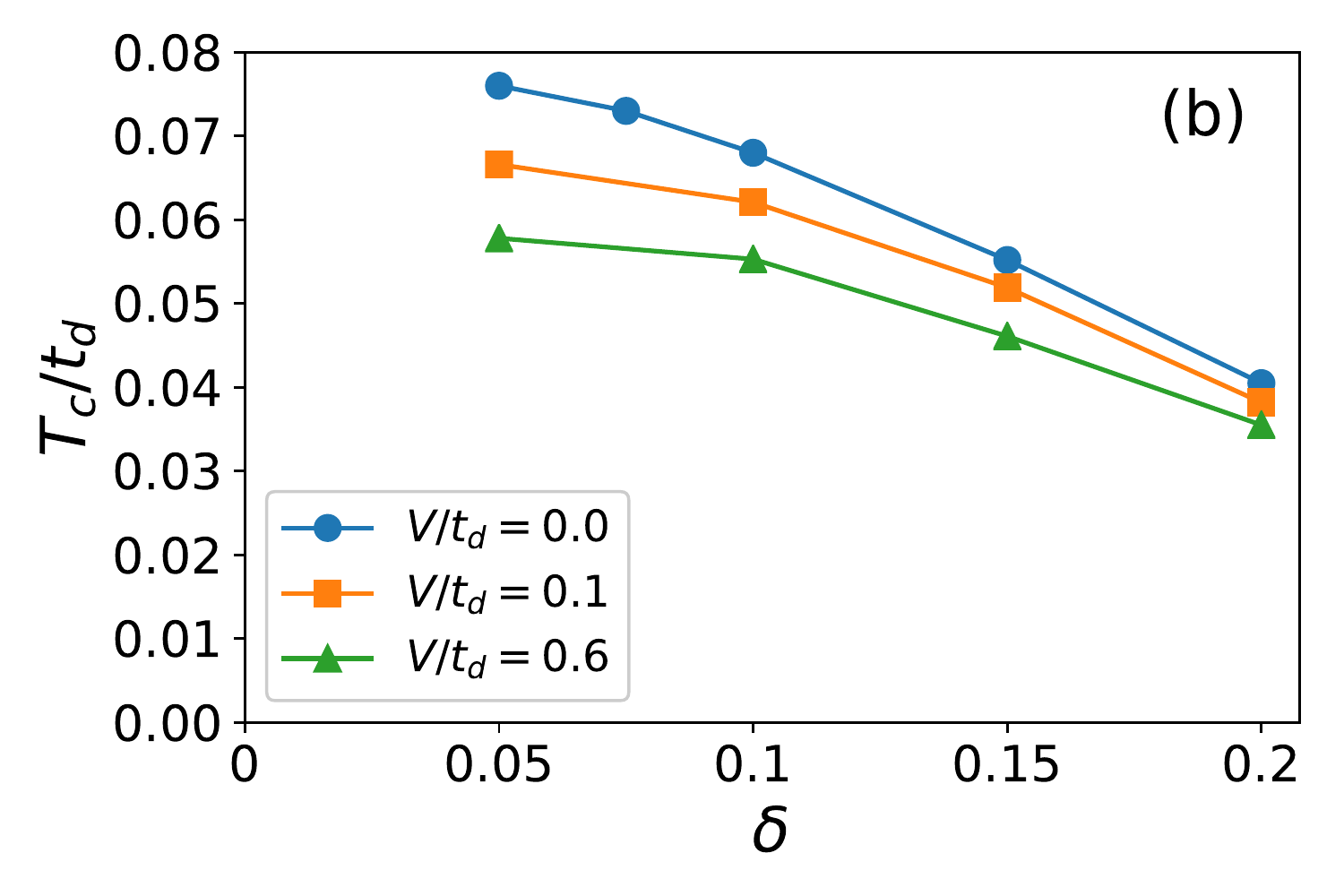,height=5.5cm,width=8.5cm,angle=0,clip} 
\psfig{figure=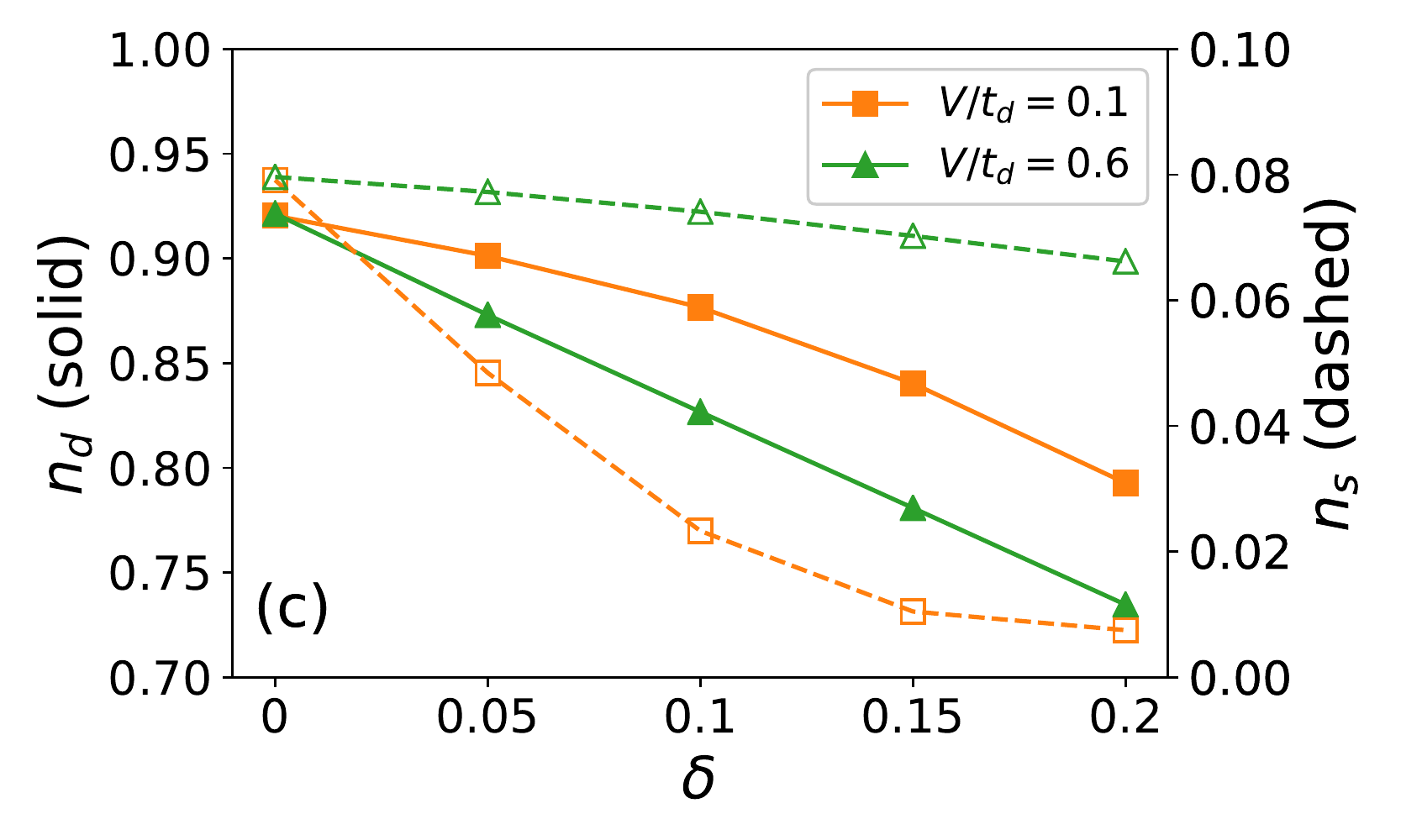,height=5.3cm,width=8.9cm,angle=0,clip} 
\caption{(a) Temperature dependence of the leading ($d_{x^2-y^2}$-wave) eigenvalue $\lambda_ {d}(T)$; (b) Doping dependence of the $d$-wave superconducting $T_{c}$ extracted from
$\lambda_d(T_c)=1$; (c) Orbital occupancy versus the doping level at $T/t_d=0.08$. The parameters are $U/t_d=7, N_c=4$.}
\label{lambda}
\end{figure}

Fig.~\ref{lambda} summarizes our key results. The panel (a) illustrates the temperature dependence of the leading $d$-wave eigenvalue $\lambda_d(T)$ for different $V$ at two fixed total electron densities $n=0.9, 0.85$, namely the total hole doping $\delta=0.1, 0.15$ respectively.
What one sees is that the presence of the hybridization suppresses $d$-wave pairing regardless of the doping level, which is expected as the additional hybridization tends to affect the original hopping integral between $d$-orbital so that the effective $U/t_d$ will be altered. In fact, even small $V/t_d=0.1$ strongly suppresses $\lambda_d(T)$ at high temperatures while this suppression is gradually diminished at lower $T$. Interestingly, at high $T$, smaller $V/t_d=0.1$ has stronger effects than larger $V/t_d=0.6$ for both $n=0.9, 0.85$. Below a crossover temperature scale $T/t_d \sim 0.2$ this difference is reversed. Although our focus here is the low temperature superconducting instability, we remark that this stronger suppression due to small $V$ above $T/t_d \sim 0.2$, which might be closely related to the promotion of $s$ orbital on SC that will be discussed later, might be worthwhile for further investigation.

Fig.~\ref{lambda}(b) compares the dependence of $T_c$ on the global doping level for the conventional Hubbard model ($V=0$) and $d$-$s$ model at two typical $V$'s. Apparently, the general feature is that the larger $V$ induces lower $T_c$ for all dopings. Nevertheless, this suppression is stronger at low dopings and it effectively diminishes at high dopings. The impact of $V$ on the variation of $T_c(\delta)$ is closely related to the evolution of $d$ and $s$ orbital's occupancies. 

As displayed in Fig.~\ref{lambda}(c), we start from the undoped system ($\delta=0$) with self-doped $d$ orbital because of finite but tiny $n_s \sim 0.08$ to be consistent with the experiments~\cite{Hall1}. At small hybridization $V/t_d=0.1$, the doped hole mainly resides on $s$ orbital as evidenced by the rapid decrease of the dashed orange curve with $\delta$. Recall that experimentally the superconductivity dome occurs in the range of $0.12<\delta<0.25$. Hence, in our relevant doping regime, namely $\delta=0.15, 0.2$, the $s$ orbital's concentration is as tiny as only $\sim 0.01$. In this sense, as pointed out by previous work~\cite{TPD_ds,Held2020,Held2022}, the $d$-$s$ model effectively reduces to the single $d$ orbital system. We numerically confirmed this single-band picture in the doping range relevant to the realistic materials. In this case, the decrease of $T_c$ shown in Fig.~\ref{lambda}(b) originates from the decrease of $d$ orbital occupancy.

\begin{figure}
\psfig{figure=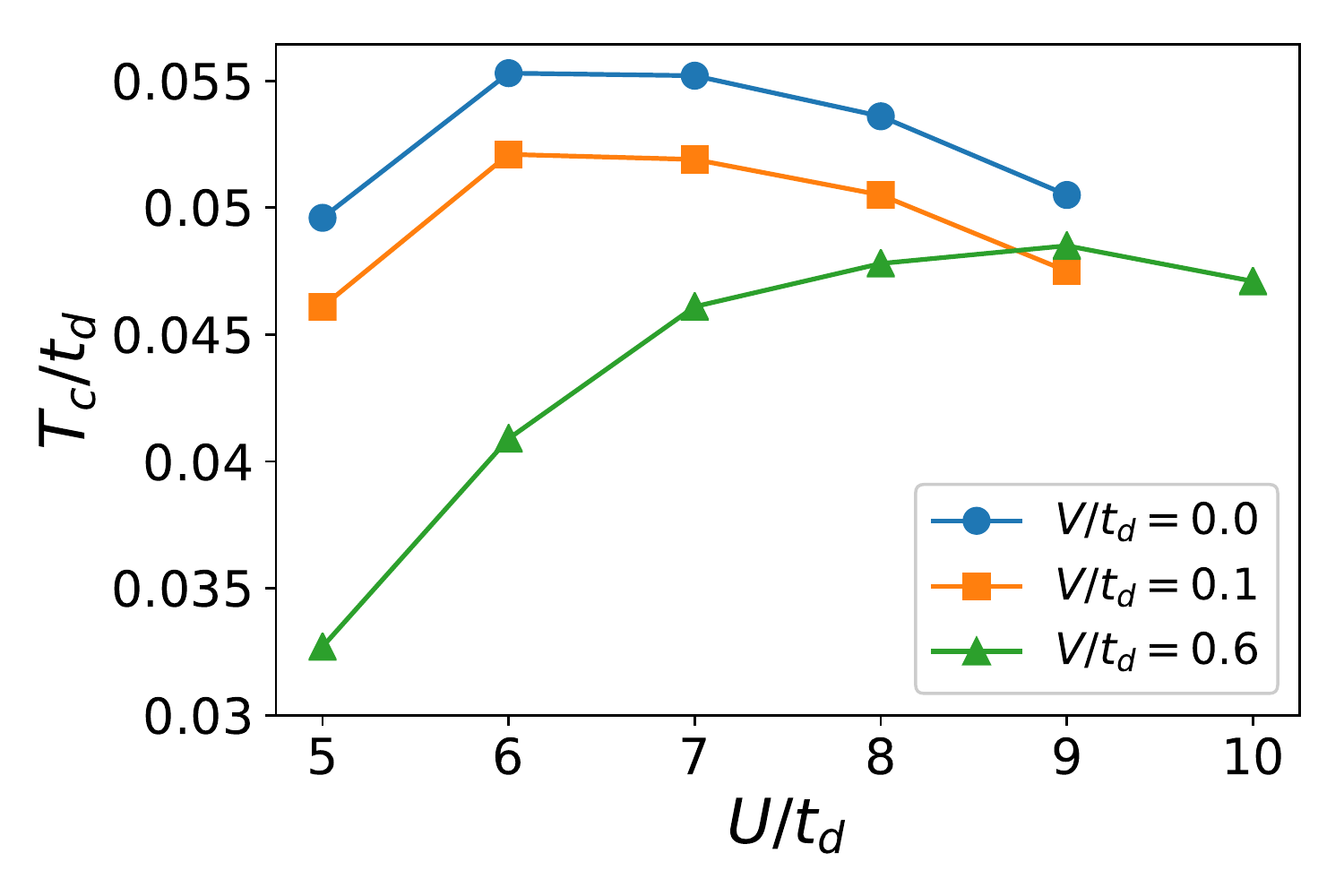,height=5.5cm,width=8.5cm,angle=0,clip} 
\caption{$T_c$ versus $U$ for $n=0.85, N_c=4$. At stronger $V/t_d=0.6$, the ``optimal'' $U$ for highest $T_c$ is shifted to a higher value.}
\label{TcU}
\end{figure}

More interestingly, at stronger $V/t_d=0.6$, both decrease of $n_d$ and $n_s$ with hole doping is smoother than the case of $V/t_d=0.1$. In fact, the doped hole dominantly goes onto $d$ orbital linearly instead, which implies that the less depleted $s$ orbital may play an important role. From the viewpoint of the single-orbital Hubbard model, larger hole doping would rapidly suppress $T_c$. Thus, in the regime of $\delta=0.15, 0.2$, it is striking to see that $T_c$ at $V/t_d=0.6$ conversely approaches to the curves for $V/t_d=0.1$ in Fig.~\ref{lambda}(b) in spite of the increasing reduce of $n_d$ for $V/t_d=0.6$ compared to $V/t_d=0.1$. The only reason originates from the plausible ``enhancing'' effect of the small but finite $n_s$ on inhibiting the suppression of SC. This argument also matches with the elevated $\lambda_d(T)$ for $V/t_d=0.6$ at high temperatures, namely the $s$ orbital's concentration tends to maintain $d$ orbital's SC. This supportive role will be more clear when we discuss the $U$-dependence of $T_c$ in Fig.~\ref{TcU}.  
In fact, the $T_c(\delta)$ for three $V$'s seemingly approaches with one another with further hole doping. Unfortunately, due to the limitation of the CT-QMC sign problem, we are not able to push to even higher doping levels and also much larger DCA cluster size to confirm this numerical phenomena to be truly physical. 

Putting another way, assuming that the electronic structure calculation~\cite{dft25} leading to $V/t_d \sim 0.6$ is more realistic than the more common belief of the smaller hybridization strength $V/t_d \sim 0.1$, the presence of the electronic density in rare-earth layer can play the vital role for supporting the superconductivity. This constitutes one of our major findings in terms of the constructive role of $d$-$s$ hybridization, which implies that a minimal model accounting for the infinite-layer nickelate SC might need consider the role of the rare-earth layer appropriately in some circumstances.

It is reasonable to argue against the supportive role of $s$ orbital on SC since the $d$-$s$ model has natural connection to PAM, where the $d$-$s$ Kondo screening has destructive effects on SC. Here we emphasize that $s$'s support on SC discussed above might only apply for the systems with tiny $s$ electron density so that the Kondo screening does not have significant impact. We also note that this extreme limit of PAM has rarely been studied before so that it is requisite to explore it in more details in future.
 
\begin{figure}
\psfig{figure=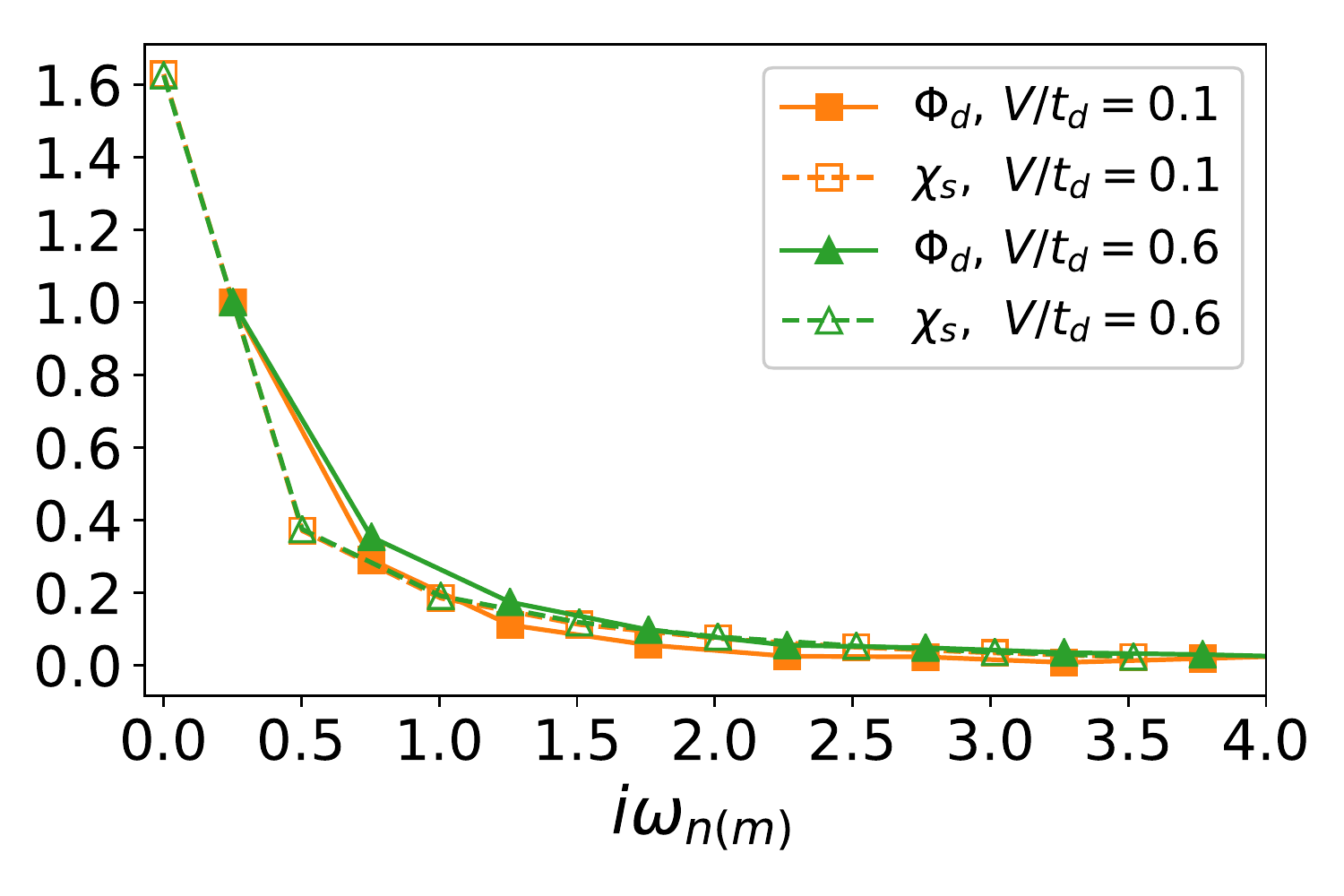,height=5.3cm,width=8.0cm,angle=0,clip}
\\ \caption{The Matsubara frequency dependence of the leading $d$-wave eigenfunction $\phi_d({\bf K},i\omega_n)/\phi_d({\bf K},i\pi T)$ with ${\bf K}=(\pi,0)$ and the normalized cluster spin susceptibility $2\chi_s(\mathbf{Q},i\omega_m)/[\chi_s(\mathbf{Q},0)+\chi_s(\mathbf{Q},i2\pi T]$ for ${\bf Q}=(\pi,\pi)$. The parameters are $U/t_d=7, N_c=4, T/t_d=0.08$ and $n=0.85$.} 
\label{eigenvector}
\end{figure}

Now that Fig.~\ref{lambda}(b) has established the notion that the impact of $V$ on the $d$-wave pairing gradually diminishes in the relatively high doping regime of $\delta=0.15-0.2$, it is natural to further explore the behavior of the pairing interaction and moreover the applicability of the spin-fluctuation mediation in $d$-$s$ model. As mentioned earlier, the structure of pairing interaction $\Gamma^{pp}$ is reflected by the BSE eigenvector $\phi_\alpha({\bf K},i\omega_n)$, which is the normal state analog of the superconducting gap function~\cite{Maier06,Scalapino06,Mi2022}.

Fig.~\ref{eigenvector} illustrates the Matsubara frequency dependence of the normalized $\phi_d({\bf K},i\omega_n)$ with ${\bf K}=(\pi,0)$. In addition, the normalized cluster spin susceptibility $\chi_s(\mathbf{Q},i\omega_m)$ for ${\bf Q}=(\pi,\pi)$, which is calculated via $\sum_{K,K'} \chi_{c}(q,K,K')$ in Eq.~\eqref{e1}, is given for comparison to check the role of spin-fluctuation in mediating the $d$-wave pairing as done in conventional Hubbard model~\cite{Maier06,Scalapino06}.
The decrease of $\phi_d$ versus $i\omega_n$ has a characteristic energy scale, which looks increasing with $V$ of a tiny amount. This indicates that the $d$-wave pairing is weakened slightly at larger $V$, which is consistent with $T_c$'s decrease in Fig.~\ref{lambda}(b).
Regarding the antiferromagnetic spin susceptibility, it has been normalized to coincide with $\phi_d(i\omega_n)$ at $\omega_n=\pi T$. Note that the Matsubara frequency entering $\chi_s$ is the bosonic $\omega_m=2m\pi T$.  
It can be seen that the antiferromagnetic spin-fluctuation spectrum is characterized by the same energy scale as the dynamics of $\phi_d(i\omega_n)$ reflecting the pairing interaction for $V/t_d=0.1$. Nevertheless, $\chi_s(i\omega_m)$ almost overlap between two cases of $V/t_d=0.1, 0.6$ and thereby the spin-fluctuation and pairing interaction have tiny discrepancy for $V/t_d=0.6$, which might be related to the non-negligible effects of $s$ orbital discuss earlier.
Fig.~\ref{eigenvector} implies that the effectively attractive interaction mediated by the antiferromagnetic fluctuations and the retardation nature of $d$-wave pairing~\cite{Scalapino06} are not qualitatively modified in the presence of $s$ orbital.

To further characterize the SC properties of $d$-$s$ model, Fig.~\ref{TcU} shows the evolution of $T_c$ versus $U$ for different hybridization $V$ at a fixed total density $n=0.85$. Apparently, the small $V$ induces a constant downshift of $T_c$ for all $U$'s, indicating that the extremely tiny concentration of $s$ orbital (shown in Fig.~\ref{lambda}(c)) do not qualitatively modify the superconducting instability. In particular, the existence of an ``optimal'' $U/t_d\sim 6$ of supporting the highest $T_c$ occurs for both single-orbital Hubbard model $V=0$ and $d$-$s$ model at $V/t_d=0.1$. This peak structure is consistent with the notion that it is important to have strong short-range antiferromagnetic correlations, which, however, decreases when $U$ becomes sufficiently large compared to the bandwidth. At strong interaction $U$, the exchange interaction $J \sim 4t^2/U$ decays so that $T_c$ decreases accordingly. 

Distinct from the weak hybridization, at relatively strong $V/t_d=0.6$, the alleviated QMC sign problem allows us to simulate the stronger $U/t_d=10$ so that the similar peak structure of $T_c(U)$ can be seen, namely the ``optimal'' $U$ is shifted to a higher value compared to the case of $V/t_d=0.1$.  Although we are not able to numerically confirm the trend at even higher $U$ for small $V$, Fig.~\ref{TcU} implies that the additional $s$ orbital and its relatively strong hybridization $V$ with $d$ orbital can have potentially supportive role for an elevated $T_c$ at strong interaction, which further backup our previous arguments discussed for Fig.~\ref{lambda}. For instance, at $U/t_d=9$, $T_c(V/t_d=0.6)>T_c(V/t_d=0.1)$. Unfortunately, it is quite challenging, if not impossible, to determine whether $T_c$ can even exceed the Hubbard model at larger $U$ with the aid of the additional $s$ orbital with appropriate hybridization $V$.

\section{Discussion}
Here we rephrase and remark on some insights on the superconducting properties of the infinite-layer nickelates and more generally other relevant correlated materials.
\begin{description}
\item [$T_c$'s magnitude:] 
Fig.~\ref{lambda}(b) reveals the pair-breaking effects of $d$-$s$ hybridization, especially at relatively strong $V/t_d=0.6$. Considering that the cuprate superconductors' optimal hole doping occurs around $\delta \sim 12.5\%$ while the infinite-layer nickelates at higher $\delta \sim 18\%$, $T_c$ has a significant drop regardless of the hybridization strength. This observation is generally consistent with the experimental fact that $T_c$ of infinite-layer nickelates is much lower than the cuprates. 
\item [Single-orbital picture:] 
The applicability of the single-orbital description depends on the hybridization $V$ strength. Fig.~\ref{lambda} indicates that if $V$ is indeed sufficiently small, it only results in a negligible decrease of $T_c$ corresponding to the tiny $s$ orbital concentration. \\
Nevertheless, if the interlayer hybridization between NiO$_2$ and R layers turns out not to be negligible, a minimal model of the infinite-layer nickelate SC or other similar compounds might need considering the role of the rare-earth layer appropriately even though the R layers have only tiny electron densities.
\item [Undoped system:]
At first glance, the curves in Fig.~\ref{lambda}(b) might hint towards a finite $T_c$ of undoped systems. This obvious discrepancy with the realistic observations probably arises from the small DCA cluster size. Previous expensive DCA calculations with larger $N_c$ but with smaller $U$ has uncovered the dome-shaped $T_c$ versus hole doping~\cite{Jarrell2013}. In this sense, our study here is simply proof-of-principle in terms of the effects of the hybridization $V$.
\item [$V$'s promotion of $T_c$:]
As discussed in Fig.~\ref{lambda} and Fig.~\ref{TcU}, it is striking to observe that the tiny but finite $s$ orbital concentration has promotion effects on SC so that the hybridization strength between NiO$_2$ and rare-earth layers might require more attention, for instance, whether the electronic structure calculation~\cite{dft25} is more realistic.
Even though this picture would be ultimately proved to be irrelevant to the infinite-layer nickelates, it might provide some insights on other strongly correlated compounds.
\item [Optimal $U$:]
Our numerical calculations uncovered some interesting trend of elevated $T_c$ due to relatively large hybridization $V$. Together with the previous point, it might provide some hints for the material design for compounds (not only infinite-layer nickelates) with higher $T_c$. 
\item [Heavy fermion physics:]
Despite that our current focus is only the superconducting properties at large hole dopings, the potential heavy fermion physics at or close to the undoped systems deserves further examination. The $d$-$s$ model has natural relation to the Anderson lattice model except for the extremely low conduction electron density. It is worthwhile exploring the potential Kondo screening exhaustion~\cite{exhaustion2019} and its impact on the magnetic properties of the undoped infinite-layer nickelates.
\item [Pairing symmetry:]
Our present work follows the established framework of $d$-wave symmetries based pairing mechanism~\cite{TPD_ds,Held2022,Held2020,Millis2022,Xianxin2020,Adhikary2020,Matianxing2020}. However, the most recent London penetration measurement strongly challenges this scenario by revealing the predominantly nodeless pairing~\cite{nodeless2022}. Specifically, there might exist a 2D-to-3D superconducting states crossover possibly due to the coupling between NiO$_2$ plane and rare-earth spacer layer. This similarity with the iron-based superconductor with nodeless multiband SC is reminiscent of the various studies on the bilayer Hubbard model at strong inter-layer hybridization, where the nodeless $s^{\pm}$-wave pairing is appreciated~\cite{bilayer2011,bilayer2016,bilayer2021}. In this regard, therefore, the $d$-$s$ model as the minimal model seems problematic at least in the regime of relatively small inter-orbital hybridization $V$. This direction may attractive further attention and exploration. 
\end{description}

\section{Summary}
In conclusion, we investigated a two-orbital $d$-$s$ model to mimic the interplay between Ni-$d_{x^2-y^2}$ and the effective interstitial s-like orbital located at the missing apical Oxygen above and below Ni in the rare-earth layers to account for the rare-earth 5d orbitals. To accurately approach to the superconducting $T_c$, we employed the dynamic cluster quantum Monte Carlo focussing on the impact of the $d$-$s$ hybridization.

It is found that the $d$-$s$ hybridization suppresses the $d$-wave pairing regardless of the doping level. The suppression of $T_c$ is stronger at low dopings and diminishes with further dopings, which is closely related to the evolution of $d$ and $s$ orbital occupancies. In particular, although the doped hole mainly resides on $s$ orbital  at small hybridization $V/t_d=0.1$ so that the $d$-$s$ model effectively reduces to the single $d$ orbital system; at stronger $V/t_d=0.6$, $d$ orbital has considerable concentration even at high doping levels, whose role is manifested by the weakened suppression trend of $V$ on SC. In other words, $T_c$ at $V/t_d=0.6$ approaches to that at $V/t_d=0.1$ at high dopings. It is thus plausible to conjecture about a striking promotion effect of the small but finite $s$ orbital occupancy on SC. 

To explore how to promote $T_c$, we further study the impact of Hubbard $U$. Although the weak hybridization $V$ does not qualitatively modify the superconducting properties without $s$ orbital, the relatively large $V/t_d=0.6$ induces that the ``optimal'' $U$ is shifted to a higher value compared to the case of $V/t_d=0.1$. This might imply a potential strategy to enhance the SC instability in general strongly correlated systems including the infinite-layer nickelates via appropriate hybridization with additional conduction electrons.

In addition, in spite of the additional $s$ orbital, the comparison between the Matsubara frequency dependence of the $d$-wave eigenvector (reflecting the pairing interaction) and the antiferromagnetic spin susceptibility supports the spin fluctuation mediation in $d$-wave pairing of $d$ orbital.

Our presented work provides complemental insights on the quantitative effects of the hybridization between NiO$_2$ and rare-earth layers on the superconducting instability relevant to the infinite-layer nickelate superconductors, especially the potentially enhancing effects of rare-earth layers on SC. As a final remark, we believe that the necessity of including the rare-earth layer in a minimal model of infinite-layer nickelates and other relevant strongly correlated compounds can be crucial in the presence of a considerable interlayer hybridization.

\section*{Acknowledgments}
We acknowledge fruitful discussion with Zhicheng Zhong and Guangming Zhang. This work was supported by National Natural Science Foundation of China (NSFC) Grant No. 12174278, the startup fund from Soochow University, and Priority Academic Program Development (PAPD) of Jiangsu Higher Education Institutions. 


\end{document}